\documentclass[aps,prl,twocolumn,showpacs,superscriptaddress]{revtex4}

\usepackage{graphicx}
\usepackage{amsmath,amssymb,bm}
\usepackage[colorlinks=true]{hyperref}

\begin{document}

\author{N.\,B.\,Kopnin } 
\affiliation{
    Low Temperature Laboratory, Aalto University, P.O. Box 15100, 00076 Aalto, Finland
}
\affiliation{
    L.\,D.\,Landau Institute for Theoretical Physics, 
    Chernogolovka, Moscow region, Akademika Semenova av., 1-A, 142432, Russia
}
\author{A.\,S.\,Mel'nikov}
\affiliation{
    Institute for Physics of Microstructures, Russian
    Academy of Sciences, 603950 Nizhny Novgorod, GSP-105, Russia
}
\author{I.\,A.\,Sadovskyy}
\affiliation{
    Materials Science Division, Argonne National Laboratory,
    9700 S. Cass Avenue, Argonne, Illinois 60637, USA
}
\author{V.\,M.\,Vinokur}
\affiliation{
    Materials Science Division, Argonne National Laboratory,
    9700 S. Cass Avenue, Argonne, Illinois 60637, USA
}

\title{Weak links in proximity superconducting two dimensional electron systems}

\begin{abstract}
We report a giant inverse proximity effect that arises in the low-dimensional devices and is crucially different from proximity in the standard superconductor-normal-superconductor (S/N/S) junctions. The normal conductors induce a giant back-action on the proximity superconducting regions even when the dimensions of the normal parts are smaller than the proximity coherence length. This essentially suppresses the superconducting characteristics of the entire system, including the critical current, well below those for the usual S/N/S structures. \end{abstract}

\pacs{
    74.45.+c,     
    74.50.+r,      
    74.78.$-$w  
}

\maketitle

A proximity effect, superconductivity, induced in a non-superconducting material brought in a contact with a superconductor, is basic to functioning of Josephson devices and weak links~\cite{Golubov04}. An explosive attention to miniature weak links is being triggered by emergent nanodevices utilizing a two-dimensional (2D) electron gas in semiconductors, graphene, semiconducting nanowires and carbon nanotubes, and topological insulators~\cite{Beenakker08,CastroNeto09,Kociak01,Charlier07,Qi11}. Practical weak link devices often have high contact resistances so that the coherence length induced by the proximity effect grows essentially longer than the coherence length of an original source superconductor. It can even exceed the dimensions of the S/N contact itself in typical devices utilizing superconducting leads and graphene sheets or nanotubes, in which the contacts are as small as tens nanometers~\cite{Heersche07a,Heersche07b,Novoselov05,Morpurgo99,Kasumov03}. 

\begin{figure}
    \includegraphics[width=0.70\linewidth]{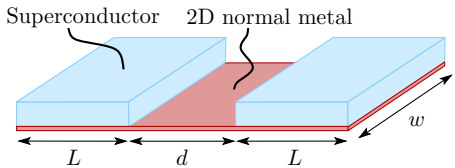}
    \caption{
        (Color online) S/2D/S structure.
        The bulk superconductors (S) make tunnel contacts with the underlying 
        2D metal regions of length~$L$ each. The are separated by a normal 
        2D weak link of length~$d$.
    }
    \label{fig:setup}
\end{figure}

We consider an S/2D/S junction of the standard experimental geometry shown in Fig.~\ref{fig:setup}. Two bulk superconducting leads (S) with the phase difference $\phi$ between them make tunnel contacts with the normal 2D metal film through an insulating layer.  The tunneling rate,~$\Gamma$, can be expressed through the barrier resistance, see below; $\Gamma$ is assumed to be small as compared to the superconducting gap $\Delta$ in the leads. The bulk superconducting leads overlap with the 2D layer in the regions $d/2<x<d/2+L$ and $-L-d/2<x<-d/2$, where~$L$ is the lead length. The width of the contact is $w$, and our system is uniform in the lateral direction. In the overlap regions the superconducting order is induced by the bulk superconducting leads due to proximity effect. For the large overlap, the induced energy gap is close to~$\Gamma$~\cite{Volkov95,Fagas05,Kopnin11}. A weak link between these proximity superconducting regions is formed by the normal part of the 2D metal of the length~$d$. We address the common experimental situation where the impurity scattering rate~$\tau^{-1}$ in disordered 2D metal film is much larger than the characteristic pairing energy scale $\Gamma$ for induced superconductivity. Our main result is the reduction of the critical current and of the induced gap by the inverse proximity effect well below the values characteristic to the usual S/N/S structures~\cite{Ambegaokar63,Kulik75a,Kulik75b}. The origin of this effect is the large value of the coherence length $\xi_{\rm\scriptscriptstyle 2D}=\sqrt{D/\Gamma}$ (here  $D=v_{\rm\scriptscriptstyle F,2D}\tau /2$ is the diffusion constant) induced in the 2D metal: Being the scale at which the superconducting coherence between the two leads forms (in conventional point contacts it forms at the weak link length $d$), it determines the superconducting characteristics of the proximity devices through the interplay with other dimensions of the contact, i.e., the weak link length~$d$ and the lead length $L$. We show [see Eqs.~(\ref{I1}) and (\ref{I2}) below] that the zero-temperature critical current is 
\begin{equation}
    I_{\rm c} \sim \epsilon_0(0)/eR_{\rm eff} \label{Ic}
\end{equation}
similarly to the standard expression for the conventional point contacts~\cite{Golubov04,Kulik75a,Kulik75b,Beenakker91} where the induced gap in the weak link for zero phase, $\epsilon_0(0)$, and the effective normal resistance of the contact, $R_{\rm eff}$, stand for the superconducting gap $\Delta$ in the leads and the weak link resistance~$R_{\rm\scriptscriptstyle N}$, respectively. However, an essential distinction from the usual expression for supercurrent is that both quantities, $\epsilon_0$ and $R_{\rm eff}$, bear strong impacts from the inverse proximity effect. We derive expressions for $\epsilon_0$ and $R_{\rm eff}$ in the most interesting situation where $d\ll \xi_{\rm\scriptscriptstyle 2D}$. 

\paragraph{Effective resistance.}
To find the contact resistance, one observes that the current conservation in the overlap region $d/2<x<L+d/2$ in the setup of Fig.~\ref{fig:setup}, with voltage $V$ across the leads, assumes the form ${\rm d}I/{\rm d}x + [V/2+\varphi(x)]/\rho_{\rm\scriptscriptstyle B}=0$, while in the normal region $0<x<d/2$, it is ${\rm d}I/{\rm d}x =0$. Here $\varphi(x)$ is the potential in the 2D layer, $-V/2$ is the potential of the right lead, and $\rho_{\rm\scriptscriptstyle B} =\hbar /4\pi \nu_2 e^2\Gamma$ is the resistance of barrier with unit area~\cite{Kopnin11},  $\nu_2=m/2\pi \hbar^2$ is the 2D density of states. Due to symmetry we consider right part of the junction, $0<x<d/2+L$. The current in the layer is $I=-\sigma_{\rm\scriptscriptstyle N} {\rm d}\varphi/{\rm d}x$ where $\sigma_{\rm\scriptscriptstyle N} = 2\nu_2 De^2$ is the normal conductivity of the 2D layer. Solving the resulting second-order differential equations for $\varphi(x)$ with the boundary conditions $\varphi (0)=0$ in the middle and ${\rm d}\varphi/{\rm d}x =0$ at the dead end $x=L+d/2$ together with continuity of the function and its derivative at $x=d/2$ one finds (see also Ref.~\onlinecite{Kupriyanov89})
\begin{equation}
    R_{\rm eff}= R_{\rm\scriptscriptstyle N} + 
    2R_{\rm\scriptscriptstyle B} (L/l)\tanh^{-1} (L/l),
    \label{R}
\end{equation}
where $R_{\rm\scriptscriptstyle B}=\rho_{\rm\scriptscriptstyle B}/Lw$ is the total normal resistance of the barrier with area $Lw$, and $l=\sqrt{\sigma_{\rm\scriptscriptstyle N} \rho_{\rm\scriptscriptstyle B}} = \xi_{\rm\scriptscriptstyle 2D}/\sqrt{2}$ is the length over which the current spreads over the whole length of the contact. At the same time, this is the spatial scale where the superconducting coherence between two leads forms. The obtained result shows that the effective contact resistance can be viewed as comprising the two resistances, the normal resistance $R_{\rm\scriptscriptstyle N}= d/\sigma_{\rm\scriptscriptstyle N} w$ and the twice the effective barrier resistance in series. If $L\gg \xi_{\rm\scriptscriptstyle 2D}$, then only parts of the barriers with length $\xi_{\rm\scriptscriptstyle 2D}$ contribute into $R_{\rm eff}$. In the opposite limit, $L\ll \xi_{\rm\scriptscriptstyle 2D}$, the full  length,~$L$, of the barrier works. Since $R_{\rm\scriptscriptstyle N}/2R_{\rm\scriptscriptstyle B}= Ld/\xi_{\rm\scriptscriptstyle 2D}^2$ the barrier resistance dominates if $Ld\ll \xi_{\rm\scriptscriptstyle 2D}^2 $, $R_{\rm eff}=2R_{\rm\scriptscriptstyle B}$. In the opposite limit, $R_{\rm eff}=R_{\rm\scriptscriptstyle N}$.

\paragraph{Induced gap and density of states.}
The second relevant characteristics of the weak link, the induced gap $\epsilon_0(\phi)$, depends on the phase difference~$\phi$. For $L\gg \xi_{\rm\scriptscriptstyle 2D}$, the induced gap $\epsilon_0(0)$ for $\phi = 0$ coincides with the homogeneous induced gap $\Gamma$. If $L\ll \xi_{\rm\scriptscriptstyle 2D}$, the superconducting coherence does not have room to develop and the induced gap drastically drops,
\begin{equation} \label{epsilon-phi1}
    \epsilon_0(\phi) = \Gamma \cos(\phi/2) / (1+d/2L).
\end{equation}
To gain a better insight into this result we note that in a long weak link with $Ld \gg \xi_{\rm\scriptscriptstyle 2D}^2 $, the barrier resistance can be neglected and the induced gap is determined by the Thouless energy, $\epsilon_0= 3.12 \epsilon_{\rm\scriptscriptstyle Th}$, where $\epsilon_{\rm\scriptscriptstyle Th}= D/d^2$ similarly to the usual ideal S/N diffusive contact~\cite{Golubov89,Pilgram00}. As the link length decreases, the barrier resistance comes into play, and the induced gap should be renormalized $\epsilon_0 \sim [R_{\rm\scriptscriptstyle N}/(R_{\rm\scriptscriptstyle N}+2R_{\rm\scriptscriptstyle B})] \epsilon_{\rm\scriptscriptstyle Th}$. For $Ld/\xi_{\rm\scriptscriptstyle 2D}^2\ll 1$ where the contact resistance is dominated by $R_{\rm\scriptscriptstyle B}$, but still $d>L$, one finds $\epsilon_0(0)=(Ld/\xi_{\rm\scriptscriptstyle 2D}^2)\epsilon_{\rm\scriptscriptstyle Th}\sim \Gamma (L/d)$ in accordance with Eq.~(\ref{epsilon-phi1}). Upon further decrease in $d$ as it becomes less than $L$, the induced gap increases further and reaches the value of the unperturbed gap $\Gamma$.

To calculate the induced gap quantitatively we develop a microscopic description based on the tunnel approximation technique of Ref.~\onlinecite{Kopnin11}. A similar method for dirty superconductors based on the quasiclassical Green function formalism and Kupriyanov-Lukichev boundary conditions~\cite{Kuprianov88} has been developed in Ref.~\onlinecite{Kupriyanov89}. The advantage of our approach over the more standard considerations~\cite{Volkov95,Fagas05,Beenakker06,Titov06,Feigelman08} is that effects of disorder and spatial variations can be easily incorporated~\cite{Kopnin13}. Our starting point are the Usadel equations for quasiclassical Green functions $g$, $f$, and $f^\dagger$ in highly disordered systems with $1/\tau \gg \Gamma$ which follow from the Eilenberger equations derived in Ref.~\onlinecite{Kopnin11}. For real frequencies, retarded (advanced) functions satisfy 
\begin{eqnarray}
    -iD_2{\bm \nabla}\left[ g{\bm \nabla}f- f{\bm \nabla}g\right]
    +2(\epsilon + \Sigma_1)f  -2\Sigma_2g =0
    \label{Useq1}
\end{eqnarray}
and similarly for $f^\dagger$ where $\Sigma_2$ is replaced with $\Sigma_2^\dagger$. Due to the normalization $g^2-ff^\dag = 1$ only two functions are independent.

The self energies $\Sigma_2=\tilde \Sigma_2e^{i\chi}$, $\Sigma_2^\dag=\tilde\Sigma_2e^{-i\chi}$ are expressed in terms of the quasiclassical Green functions of the bulk superconductor. For $-|\Delta|< \epsilon <|\Delta|$, where $|\Delta|$ is the bulk superconducting gap,
\begin{equation}
    \Sigma_1 =
        \Gamma \epsilon/\sqrt{|\Delta|^2-\epsilon^2}, \quad
    \tilde\Sigma_2 =
        \Gamma |\Delta|/\sqrt{|\Delta|^2-\epsilon^2}
    \label{Sigmas}
\end{equation}
in the overlap regions $d/2<|x|<L+d/2$ are independent of coordinates. For $|\epsilon|>|\Delta|$ the self energies have to be analytically continued onto the upper or lower half-plane of complex $\epsilon$ for retarded or advanced functions, respectively. In what follows we consider low tunneling rate $\Gamma \ll \Delta$. For $\Gamma \ll |\Delta|$ and $\epsilon \ll \Delta$ we have $\tilde \Sigma_2 = \Gamma$, $\Sigma_1 \ll \epsilon$. We assume $\chi =\phi/2$ for $x>d/2$ and $\chi =-\phi/2$ for $x<-d/2$. Inside the normal 2D layer, $|x|<d/2$ the self energies vanish, $\Sigma_1=\tilde \Sigma_2=0$.

The Green function for a homogeneous proximity system is $g^2 = \epsilon ^2/(\epsilon ^2-\Gamma^2) $ such that the induced gap is equal to $\Gamma$. In our paper we are interested in temperatures much lower than $\Gamma$ and in energies of the order or lower than $\Gamma$. This is the energy scale where the inverse proximity effect is most pronounced.

Finite sizes of the leads and of the weak link region result in a decrease in the induced gap as compared to its uniform value $\Gamma$. Let us consider first zero phase difference $\phi =0$ and assume that the induced gap $\epsilon_0$ is smaller than $\Gamma$. We look for a solution that corresponds to energies below the induced gap. In this case the retarded and advanced functions coincide, they are imaginary and can be written in the form $g=-i\sinh \theta$ and $f=f^\dag =-i\cosh \theta$ with real $\theta>0$ for $0<\epsilon <\epsilon_0$.  The Usadel equations 
have the first integrals. For $d/2<x<L+d/2$
\begin{equation}
    \int_{\theta_L}^{\theta_d} 
    \frac{ {\rm d}\theta}{\left[ \cosh(\theta - \theta_\infty) - \cosh(\theta_L - \theta_\infty)\right]^{1/2}} 
    =\sqrt{\frac{4\Sigma_\epsilon L^2 }{D}},
    \quad \label{theta+}
\end{equation}
where $\Sigma^2_\epsilon = \Gamma^2 - \epsilon^2$ and $\sinh \theta_\infty =  \epsilon /\Sigma_\epsilon.$ The integration constant is chosen to satisfy the boundary conditions ${\rm d}\theta/{\rm d}x =0, \; \theta =\theta_L$ at $x=L+d/2$, while $\theta_d$ is the angle at $x=d/2$. The angles satisfy $\theta_d>\theta_L>\theta_\infty$. In the region $|x|<d/2$ we have 
\begin{equation}
    \int_{\theta_d}^{\theta_0} \frac{{\rm d}\theta}{\left[ \sinh \theta_0 -\sinh \theta\right]^{1/2}} = 
    \sqrt{\frac{\epsilon d^2}{D}},
    \label{theta-}
\end{equation}
where ${\rm d}\theta/{\rm d}x =0,$ $\theta =\theta_0$ at $x=0$. Continuity of derivatives at $x=d/2$ yields
\begin{multline}
    \epsilon \left[ \sinh \theta_0 -\sinh \theta_d\right] \\
    = \Sigma_\epsilon 
        \left[ \cosh(\theta_d\! - \! \theta_\infty) 
        - \cosh(\theta_L\! - \! \theta_\infty) \right].
    \label{bcond-dL}
\end{multline}
Equations~(\ref{theta+}), (\ref{theta-}), and (\ref{bcond-dL}) allow one to find $\theta_d$, $\theta_L$, and $\theta_0$ together with the Green function in each region.

For long leads, $L\gg \xi_{\rm\scriptscriptstyle 2D}$, the angles should satisfy $\theta_L=\theta_\infty$. For $\epsilon \ll \Sigma_2$ one has $\theta_\infty=0$. Eq.~(\ref{bcond-dL}) then requires $\theta_d =\theta_\infty =0$. With $\theta_d=0$ Eq.~(\ref{theta-}) determines the induced gap which coincides with the usual Thouless gap for diffusive SN junctions~\cite{Golubov89,Pilgram00}, 
\begin{equation}
    \epsilon_0 = 3.12 D/d^2.
    \label{epsilon_Th}
\end{equation}
For $\epsilon >\epsilon_0$ the real solution for $\theta$ does not exist, thus $g^{\rm\scriptscriptstyle R} \ne g^{\rm\scriptscriptstyle A}$, leading to a finite density of states. This result holds as long as $D/d^2 \ll \Gamma$ or $d\gg \xi_{\rm\scriptscriptstyle 2D}$.

\begin{figure}[tb]
    \includegraphics[width=0.90\linewidth]{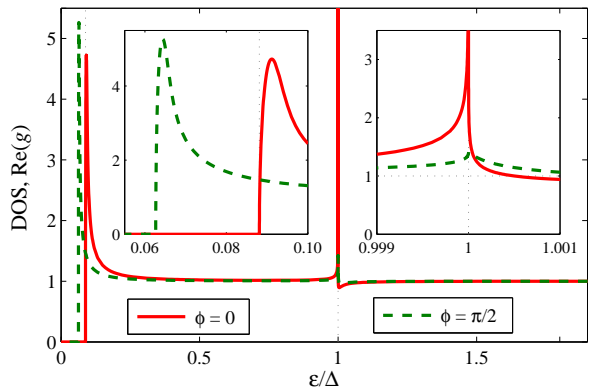}
    \caption{
        (Color online) DOS at $x=0$ as a function of energy. There are two 
        singularities in the energy spectrum: at superconducting gap $\Delta$ 
        and at induced gap $\epsilon_0$ as shown in corresponding insets. 
        DOS vanishes below~$\epsilon_0$. Here $\Gamma/|\Delta| = 0.2$, 
        $d/2L = 1$, and $\xi_{\rm\scriptscriptstyle 2D}/L = 8.94$. 
    }
    \label{fig:DOS}
\end{figure}

For short $L\ll \xi_{\rm\scriptscriptstyle 2D}$ one has $\theta_d\approx \theta_L$. Expanding Eqs.~(\ref{theta+}) and (\ref{bcond-dL}) in small $\theta_d-\theta_L$ we find from Eq.~(\ref{theta-})
\begin{equation}
    \int_{\theta_d}^{\theta_0} \frac{{\rm d}\theta}{[\sinh \theta_0 -\sinh \theta]^{1/2}} =
    \frac{\Sigma_\epsilon Ld}{D} \frac{\sinh(\theta_d -\theta_\infty)}{[\sinh \theta_0 -\sinh \theta_d]^{1/2}}.
    \label{eq-bc}
\end{equation}
This equation shows that $\Sigma_\epsilon Ld/D\sim R_{\rm\scriptscriptstyle N}/R_{\rm\scriptscriptstyle B}$ is the important parameter that governs the behavior of contacts with short leads, $L\ll \xi_{\rm\scriptscriptstyle 2D}$. For $Ld/\xi_{\rm\scriptscriptstyle 2D}^2 \ll 1$ the angle $\theta_0$ is close to $\theta_d$. Expanding Eq.~(\ref{eq-bc}) in small $\theta_0-\theta_d$ yields
\begin{equation}
    \epsilon  = \epsilon_0 \tanh \theta_d, \quad
    \epsilon_0= \Gamma/(1+d/2L)
    \label{epsilon0}
\end{equation}
and
$
    g^{\rm\scriptscriptstyle R(A)} = 
    -i \sinh \theta_d = 
    -i \epsilon/\sqrt{\epsilon_0^2 - \epsilon^2}.
$
A solution of Eq.~(\ref{epsilon0}) exists if $\epsilon \leqslant \epsilon_0$. Therefore, $\epsilon_0$ is the induced gap in the both normal and superconducting regions. One can show that the phase difference $\phi$ between the superconducting leads results in further reduction of the induced gap according to Eq.~(\ref{epsilon-phi1}).

\begin{figure}[tb]
    \includegraphics[width=0.90\linewidth]{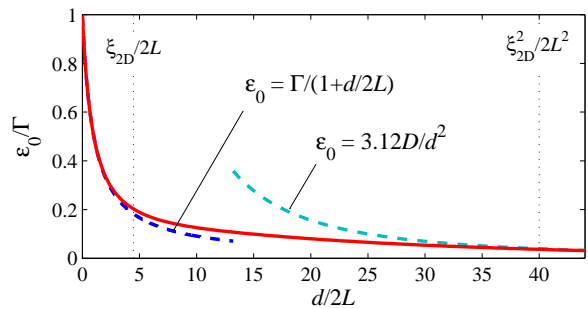}
    \caption{
        (Color online) The induced gap $\epsilon_0$ as a function of $d$ 
        for the same parameters as in Fig.~\ref{fig:DOS} and $\phi = 0$.
        The dashed lines show limiting cases given by Eq.~(\ref{epsilon0}) 
        and Thouless gap Eq.~(\ref{epsilon_Th}) for diffusive SN junctions 
        correspondingly.
    }
    \label{fig:gap}
\end{figure}

If $L\ll \xi_{\rm\scriptscriptstyle 2D}$ while $d$ is sufficiently long, $Ld \gg \xi_{\rm\scriptscriptstyle 2D}^2$,  Eq.~(\ref{eq-bc})  implies $\theta_d \to \theta_\infty$ where $\theta_\infty = 0$ since $\epsilon \ll \Gamma$. Eq.~(\ref{theta-}) with $\theta_d =0$
leads to the Thouless gap. Therefore, for $L\ll \xi_{\rm\scriptscriptstyle 2D}$, the induced gap in the normal region decreases below $\Gamma$ with increasing $d$ first as Eq.~(\ref{epsilon0}) for $d \ll \xi_{\rm\scriptscriptstyle 2D}^2/L$ and then as $d^{-2}$ for $d\gg \xi_{\rm\scriptscriptstyle 2D}^2/L$. 

We calculated the induced energy gap and the density of states (DOS). Shown in Figs.~\ref{fig:DOS} and \ref{fig:gap} are the DOS in the normal region as a function of energy and the induced gap for various values of~$d$ obtained by numerical solution of the Usadel equations (\ref{Useq1}). 

Our analysis confirms the earlier statement that the induced gap in a junction comprising regions with induced superconductivity considerably differs from that in the usual SNS or SINIS junctions: It is modified through the interplay between the additional length scale $\xi_{\rm\scriptscriptstyle 2D}$ and other dimensions of the system. For example, if both, the contact size $L$ and the weak link length $d$ are longer than $\xi_{\rm\scriptscriptstyle 2D}$, the induced gap approaches the usual Thouless gap Eq.~(\ref{epsilon_Th}). In this limit, properties of the junction can perhaps be described by a model of Ref.~\onlinecite{Feigelman08} where the Usadel equations with boundary conditions of the Kupriyanov-Lukichev type with an effective interface conductance $G$ were used at the circumference of a superconducting disk placed on top of a 2D layer (graphene). However, for shorter contact dimensions, when $L,d\ll \xi_{\rm\scriptscriptstyle 2D}$, the induced gap Eq.~(\ref{epsilon0}) decreases much below the Thouless gap for given $d$ (see Fig.~\ref{fig:gap}) due to the giant inverse proximity effect. In this limit, our results drastically differ from those of Ref.~\onlinecite{Feigelman08} where the length scale $\xi_{\rm\scriptscriptstyle 2D}$ is absent. Note that simultaneously in short contacts the effective conductance $G$ becomes a function of system dimensions and, possibly, also of energy rather than merely being a material constant. A comment on the boundary condition, which employs an approximation of the effective energy-independent induced ``order parameter''~\cite{Beenakker06} used to model the induced superconductivity,  is in order. One sees from Eq.~(\ref{Sigmas}) that this assumption is of rather limited validity and holds only for a spatially uniform case and for low barrier transparency $\Gamma \ll \Delta$ where the ``order parameter'' does not depend on the energy gap $\Delta$ in the bulk superconductor.

\paragraph{Supercurrent.}
We calculate the supercurrent for short weak links, $d\ll \xi_{\rm\scriptscriptstyle 2D}$, assuming for simplicity that the Usadel equations can be linearized in small $f$ and $f^\dagger$. We shall see that this approximation gives the correct result for most of the situations of interest. In Matsubara representation one has for $d/2<|x|<L+d/2$
\begin{equation*}
    l_\omega^2 \frac{{\rm d}^2 f}{{\rm d}x^2} = 
    f -\frac{i\Gamma }{\omega_n}e^{{\rm sign}(x)i\phi/2},
\end{equation*}
where $l_\omega ^2 = D/2\omega_n$ and $\omega_n =2\pi T(n+1/2)$ is the Matsubara frequency. For $|x|<d/2$ one has the same equation but with $\Gamma =0$. Equations for $f^\dagger$ are obtained by replacing $\phi \to -\phi$. The boundary conditions are ${\rm d}f/{\rm d}x =0$ at $x=\pm (L+d/2)$ and continuity of $f$ and ${\rm d}f/{\rm d}x$ at $x=d/2$. Solving these equations we find for $|x|<d/2$
\begin{align}
    f(x)= \frac{i\Gamma }{\omega_n}\cos(\phi/2)
    \frac{\sinh(L/l_\omega)\cosh(x/l_\omega)}{\sinh[(L+d/2)/l_\omega]} \nonumber \\
    - \frac{\Gamma }{\omega_n}\sin(\phi/2)
    \frac{\sinh(L/l_\omega)\sinh(x/l_\omega)}{\cosh[(L+d/2)/l_\omega]}
\end{align}
and similarly for $f^\dagger$ with $\phi \to - \phi$. The linear approximation holds as long as $f\ll 1$, i.e., $\omega \gg \omega_{\rm min}$ where
\begin{equation}
    \omega_{\rm min}= \Gamma \sinh(L/l_\omega)/\sinh[(L+d/2)/l_\omega].
    \label{lin-cond}
\end{equation}
The supercurrent has the form
\begin{eqnarray}
    I=\frac{d}{2eR_{\rm\scriptscriptstyle N}}\pi i T
    \sum_n \left[ f^\dagger \frac{{\rm d}f}{{\rm d}x}-f\frac{{\rm d}f^\dagger}{{\rm d}x}\right].
\end{eqnarray}
Calculating current at $x=0$ we find $I=I_{\rm c} \sin \phi$ where
\begin{eqnarray}
    I_{\rm c} = \frac{\Gamma^2d}{eR_{\rm\scriptscriptstyle N}} 2\pi T
    \sum_{n>0} \frac{\sinh^2(L/l_\omega)}{\omega_n^2 l_\omega \sinh[(2L+d)/l_\omega]}.
    \label{Ic-lin}
\end{eqnarray}
The sum converges when $L\gg l_\omega$ yielding 
\begin{equation}
    I_{\rm c} \sim (\Gamma/eR_{\rm\scriptscriptstyle N})(d/\xi_{\rm\scriptscriptstyle 2D}). 
    \label{I1}
\end{equation}
This expression agrees with Eq.~(\ref{Ic}) since $\epsilon_0 =\Gamma$ and $R_{\rm eff}= R_{\rm\scriptscriptstyle N}(\sqrt{2}\xi_{\rm\scriptscriptstyle 2D}/d)$ for $L\gg \xi_{\rm\scriptscriptstyle 2D}$ and $d\ll \xi_{\rm\scriptscriptstyle 2D}$. However, this estimate holds only by the order of magnitude because the linear approximation does not work for frequencies $\omega \sim \Gamma$ [see Eq.~(\ref{lin-cond})] which give the main contribution to the sum. Nevertheless, Eq.~(\ref{Ic-lin}) can be used exactly for $L, d\ll l_\omega$ when the sum diverges logarithmically and thus it is determined by larger frequencies at which linear approximation holds. In this case
\begin{eqnarray}
    I_{\rm c} = \frac{d\Gamma^2}{eR_{\rm\scriptscriptstyle N}D}\frac{L^2}{L+d/2} \sum_{n>0} \frac{2\pi T}{\omega_n}
    = \frac{\epsilon_0 }{eR_{\rm eff}} \log \frac{\omega_{\rm max}}{\omega_{\rm min}},
    \label{I2}
\end{eqnarray}
where $\epsilon_0$ is given by Eq.~(\ref{epsilon0}). This expression has the form of Eq.~(\ref{Ic}) from the introduction. The upper cutoff is set by the condition $L+d/2 \sim l_\omega$ when the sum in Eq.~(\ref{Ic-lin}) starts to converge; i.e., $\omega_{\rm max}= \Gamma \xi_{\rm\scriptscriptstyle 2D}^2/(L+d/2)^2$. For $L,d \ll l_\omega$ we find from Eq.~(\ref{lin-cond}) $\omega_{\rm min}=\epsilon_0$. Therefore, 
\begin{eqnarray*}
    \omega_{\rm max}/\omega_{\rm min} = 
    \xi_{\rm\scriptscriptstyle 2D}^2/L(L+d/2). 
\end{eqnarray*}
Equation (\ref{I2}) holds when this ratio is large. For $d\ll L$ we need $L\ll \xi_{\rm\scriptscriptstyle 2D}$ while the limit $d\gg L$ requires $Ld \ll \xi_{\rm\scriptscriptstyle 2D}^2$. In both cases the effective resistance $R_{\rm eff}=2R_{\rm\scriptscriptstyle B}$. 

\paragraph{Summary.} 
The S/2D/S junctions, in which two regions with proximity induced superconductivity are coupled through a 2D normal-metal layer, differ considerably from the usual S/N/S or S/I/N/I/S structures. This difference stems from the fact that superconductivity induced in the proximity regions linked by tunnel contacts with the bulk superconducting leads is weak and the corresponding coherence length $\xi_{\rm\scriptscriptstyle 2D}$ is much longer than that in the bulk superconductors. For contact dimensions smaller than $\xi_{\rm\scriptscriptstyle 2D}$, the inverse proximity decreases dramatically the induced gap in the 2D layer and the critical current of the junction acquires strong dependences upon the weak-link length $d$ and the lead length $L$.

\acknowledgements
This work was supported by the U.S. Department of Energy Office of Science under the Contract No. DEAC02-06CH11357. The work of NK was also supported by the Academy of Finland through its Centre of Excellence Program (Project no.~250280) and by the European Union Seventh Framework Programme (FP7/2007-2013) under grant agreement no.~308850.


\begin{thebibliography}{99}

\bibitem{Golubov04}
   A.A. Golubov, M.Yu. Kupriyanov, and E. Il'ichev, 
    Rev. Mod. Phys. {\bf 76}, 411--469 (2004).

\bibitem{Beenakker08}
    C.W.J. Beenakker,
    Rev. Mod. Phys. {\bf 80}, 1337--1354, (2008).

\bibitem{CastroNeto09}
    A.H. Castro Neto, F. Guinea, N.M. Peres, K.S. Novoselov, and A.K. Geim, 
    Rev. Mod. Phys. {\bf 81}, 109--162 (2009).

\bibitem{Kociak01}
    M. Kociak, A.Yu. Kasumov, S. Gu{\'e}ron, B. Reulet, I.I. Khodos, Yu.B. Gorbatov, V.T. Volkov, L. Vaccarini, and H. Bouchiat,
    Phys. Rev. Lett. {\bf 86}, 2416--2419 (2001).

\bibitem{Charlier07}
    J.C. Charlier, X. Blase, and S. Roche,
    Rev. Mod. Phys. {\bf 79}, 677--732 (2007).

\bibitem{Qi11}
    X.-L. Qi and S.-C. Zhang,
    Rev. Mod. Phys. {\bf 83}, 1057--1110 (2011).

\bibitem{Heersche07a}
    H.B. Heersche, P. Jarillo-Herrero, J.B. Oostinga, L.M.K. Vandersypen, and A.F. Morpurgo,
    Solid State Commun. {\bf 143}, 72--76 (2007).

\bibitem{Heersche07b}
    H.B. Heersche, P. Jarillo-Herrero, J.B. Oostinga, L.M.K. Vandersypen, and A.F. Morpurgo,
    Nature {\bf 446}, 56--59 (2007).

\bibitem{Novoselov05}
    K.S. Novoselov, A.K. Geim, S.V. Morozov, D. Jiang, M.I. Katsnelson, I.V. Grigorieva, S.V. Dubonos, and A.A. Firsov,        
    Nature {\bf 438}, 197--200 (2005).

\bibitem{Morpurgo99}
    A.F. Morpurgo, J. Kong, C.M. Marcus, and H. Dai,
    Science {\bf 286}, 263--265 (1999).

\bibitem{Kasumov03}
    A. Kasumov, M. Kociak, M. Ferrier, R. Deblock, S. Gu{\'e}ron, B. Reulet, I. Khodos, O. St{\'e}phan, and H. Bouchiat,
    Phys. Rev. B {\bf 68}, 214521 (2003).
    
\bibitem{Volkov95}
    A.F. Volkov, P.H.C. Magn{\' e}e, B.J. van Wees, and T.M. Klapwijk,
    Physica C {\bf 242}, 261--266 (1995).

\bibitem{Fagas05}
    G. Fagas, G. Tkachov, A. Pfund, and K. Richter,
    Phys. Rev. B {\bf 71}, 224510 (2005).

\bibitem{Kopnin11}
    N.B. Kopnin and A.S. Mel'nikov, 
    Phys. Rev. B {\bf 84}, 064524 (2011).

\bibitem{Ambegaokar63}
    V. Ambegaokar and A. Baratoff,
    Phys. Rev. Lett. {\bf 10}, 486--489 (1963); 
    Errata,
    {\bf 11} 104--104 (1963).

\bibitem{Kulik75a}
    I.O. Kulik and A.N. Omel'yanchuk,
    Pis'ma Zh. Eksp. Teor. Fiz. 21, 216 (1975)
    [JETP Lett. {\bf 21}, 96 (1975)]. 

\bibitem{Kulik75b}
    I.O. Kulik and A.N. Omel'yanchuk,
    Fiz. Nizk. Temp. {\bf 4}, 296--311 (1978).
    [Sov. J. Low Temp. Phys. {\bf 4}, 142 (1978)].

\bibitem{Beenakker91} 
    C.W.J. Beenakker, 
    Phys. Rev. Lett. {\bf 67}, 3836 (1991).
    
\bibitem{Kupriyanov89}
    M.Yu. Kupriyanov,
    Sverhprovodimost': Fizika, Khimia, Tekhnika (Sov. Journ. Superconductivity) {\bf 2}, 5 (1989).    

\bibitem{Golubov89}
    A.A. Golubov and M.Yu. Kupriyanov,
    Zh. Eksp. Teor. Fiz. {\bf 96}, 1420--1433 (1989).
    [Sov. Phys. JETP {\bf 69} 805 (1989)].

\bibitem{Pilgram00}
    S. Pilgram, W. Belzig, and C. Bruder,
    Phys. Rev. B {\bf 62}, 12462--12467 (2000).  

\bibitem{Kuprianov88}
    M.Yu. Kuprianov and V.F. Lukichev,
    Zh. Eksp. Teor. Fiz.  {\bf 94}, 139--149 (1988) 
    [Sov. Phys. JETP {\bf 67}, 1163 (1988)].

\bibitem{Beenakker06}
    C.W.J. Beenakker,
    Phys. Rev. Lett. {\bf 97}, 067007 (2006).

\bibitem{Titov06}
    M. Titov and C.W.J. Beenakker,
    Phys. Rev. B {\bf 74}, 041401(R) (2006).

\bibitem{Feigelman08}
    M.V. Feigel'man, M.A. Skvortsov, and K.S. Tikhonov,
    Pis'ma ZhETF {\bf 88}, 862 (2008)
    [JETP Lett.  {\bf 88}, 747 (2008)].

\bibitem{Kopnin13}
    N.B. Kopnin, I.M. Khaymovich, and A.S. Mel'nikov,
    Phys. Rev. Lett. {\bf 110}, 027003 (2013).

\end{thebibliography}
\end{document}